\documentclass[pop,fleqn,a4paper]{w-art}
\usepackage{amsmath,amsfonts,amssymb,times,w-thm,graphicx,stmaryrd}

\def\be#1#2\ee{\begin{equation}\label{#1}#2\end{equation}}
\def\ben#1\een{\begin{equation}\nonumber #1\end{equation}}
\def\bea#1#2\eea{\begin{eqnarray}\label{#1}#2\end{eqnarray}}
\def\note#1{\marginpar{\raggedright\if@twoside\ifodd\c@page\raggedleft\fi\fi\sf\scriptsize Note: #1}}

\newcommand{\TT}{T\oplus T^{\ast}}
\def\vol{\mathop{vol}\nolimits}
\def\Re{\mathop{\rm Re}\nolimits}

\def\spin{\mathop{\rm Spin}\nolimits}
\newcommand{\eunorm}[1]{\parallel\mbox{\hspace{-3pt}} #1\mbox{\hspace{-3pt}}\parallel}
\newcommand{\mf}{\mathfrak}
\newcommand{\GxG}[1]{#1\!\times\!#1}
\newcommand{\GG}{\GxG{G}}

\begin{document}
\pagespan{1}{}
\bigskip

\begin{flushright}
NIKHEF/2006-012\\
\bigskip
\end{flushright}


\title[Calibrations on spaces with GxG--structure]{Calibrations on spaces with $\mathbf{\GG}$--structure}

\author[F. Gmeiner]{Florian Gmeiner\inst{1}%
}
\author[F. Witt]{Frederik Witt\inst{2}%
}

\address[\inst{1}]{NIKHEF, Kruislaan 409, 1098 SJ Amsterdam, The Netherlands}
\address[\inst{2}]{Freie Universit\"at Berlin, Arminallee 3, 14195 Berlin, Germany}

\begin{abstract}
In these notes we give an introduction to the concept of spaces with $\GG$--structure and their structured submanifolds. These objects generalise the classical notion of a calibrated submanifold. Therefore, they are interesting from a string theory viewpoint as they are relevant to describe D--branes in string compactifications on backgrounds with fluxes.

\bigskip\bigskip
\noindent\emph{This article is based on a talk given by the first author at the RTN Workshop ''Constituents, Fundamental Forces and Symmetries of the Universe'', Naples, Italy, 9-13 Oct 2006.}
\end{abstract}

\maketitle

\section{Introduction}
Compactifications of type II string theory with D--branes are very interesting from the point of view of string phenomenology and have been studied extensively in recent years. Generically, the effective four dimensional theories obtained this way contain several massless scalars, the so-called moduli. In order to obtain a sensible theory, one would like to have a potential to fix their values.
This can be achieved by introducing background fluxes, non--trivial values for the NS--NS and R--R field strength in the six--dimensional manifold used to compactify the theory (for a recent review on type II compactifications with and without fluxes see~\cite{flrev}).
Introducing general fluxes, one leaves the realm of Calabi--Yau compactifications and deals with backgrounds with an $\GxG{SU(3)}$--structure~\cite{jewi,glw}.

These geometries are a special case of $\GG$--structures, introduced in~\cite{wi05}, and based on the notion of so--called generalised geometry, which goes back to Hitchin~\cite{hi02}.
Possible applications in string theory are not limited to compactifications of type II, but might also be relevant to compactifications of M--theory~\cite{ts06}.

Introducing higher--dimensional objects, such as D--branes in type II, in spaces of generalised geometry triggers new questions,
in particular on the amount of preserved supersymmetry in the four--dimensional theory. In the classical case of $SU(3)$--compactifications, this question can be answered by considering calibrations~\cite{hala82}.
As it turns out, the concept of a calibration also makes sense in spaces with $\GG$--structure~\cite{gmwi06}.
Moreover, this notion of a generalised calibration includes the classical case as a special one.
In the context of D--branes in type II string theory, such calibrated submanifolds have been considered in~\cite{koma}.
The topic is subject to active research and many aspects are still unexplored, in particular we miss explicit realisations of this concept in string theory. Recent developments~\cite{morekoma} include the investigation of the superpotential of compactified type II theories on $\GxG{SU(3)}$--spaces with D--branes and deformations of the calibrated cycles (cf. also the contribution of L. Martucci to these proceedings).

This article is organised as follows. First we will explain the notion of spaces with $\GG$--structure, paving the ground for our main topic, the notion of calibrated submanifolds on these spaces which we discuss in Section~\ref{secCal}. Finally, an explicit example showing how to obtain the classical calibrations will be discussed in the case of an $\GxG{SU(3)}$--structure.

\section[GxG--structures]{$\mathbf{\GG}$--structures}\label{secGG}
Before defining $\GG$--structures, let us briefly review the notion of a ``classical'' $G$--structure and what we mean by topological and geometrical reductions\footnote{For a mathematically more sophisticated treatment of the content of this section see e.~g.~\cite{wi06}.}. Given a manifold $M$ of dimension $n$ and a vector bundle $V$ of rank $m$ over it, the transition functions generically take values in $GL(m)$. A \emph{topological reduction} or {\em $G$--structure} is defined by requiring the transition functions to take values in $G\subset GL(m)$. Objects $Q_i$, which are stabilised by the action of $G$, i. e. they are invariant under $G$--transformations, can therefore be globally defined on $M$. The possible choices for embedding $G$ into $GL(m)$ are parametrised by $GL(m)/G$. This also gives rise to topological obstruction against the existence of a $G$--structure, namely we need to have at least one globally defined section of the associated bundle with fibre $GL(m)/G$. In the following we will also speak of a {\em reduction} of the structure group $GL(m)$ to $G$. A $G$--structure on the tangent bundle $T$ is refered to as {\em classical}.

As a simple example consider the case of a manifold of dimension $n$ with an $SO(n)$--structure. The reduction of the structure group from $GL(n)$ to $O(n)$ stabilises a symmetric two--tensor, which we identify as the Riemannian metric $g$. A further reduction from $O(n)$ to $SO(n)$ is equivalent to the existence of a globally defined volume form, that is an orientation on $M$. Here we find a topological obstruction against the existence of an $SO(n)$--structure, namely $M$ has to be orientable, which is equivalent to $w_1(M)=0$, the vanishing of the first Stiefel--Whitney class. On $M$ exists a prefered connection, the Levi-Cevita connection, which is compatible with the $SO(n)$--structure, meaning $\nabla^{LC} g=0$. This links into the concept of a \emph{geometrical reduction}. An arbitrary connection $\nabla$ on $T$ can be locally written as $d+\omega$, where $d$ is usual differentiation on $\mathbb{R}^n$ and $\omega$ a $1$--form taking values in the Lie algebra $\mf{gl}(n)$. A topological reduction to $G$ is compatible with $\nabla$ if the locally defined matrices actually take values in $\mf{g}$, the Lie algebra of $G$. This is equivalent to the statement that the stabilised objects $Q_i$ are covariantly constant, i.~e. $\nabla Q_i=0$, as it happens, for instance, for $g$ and the Levi--Civita connection in the case of an $O(n)$--structure.

On our way to $\GG$--structures we will take two more intermediate steps. Firstly, we consider so--called \emph{generalised structures}, which are associated with a subgroup of $SO(n,n)$. This is the structure group of the bundle $E=\TT$, which naturally exists on any manifold $M$. It is a rank $2n$ vector bundle with a natural orientation and inner product of signature $(n,n)$. The $SO(n,n)$--structure on $E$ is always spinnable, for $w_2(E)=w_2(T)+w_1(T)\cup w_1(T^*)+w_2(T^*)=0$ (using $w_1(T^*)=w_1(T)$ and $w_2(T^*)=w_2(T)$), hence we obtain a corresponding spin structure (which is actually canonic). The corresponding spinors, sections of the spin bundles $S_\pm$ of $\spin(n,n)$, are almost isomorphic to even or odd forms on $M$. Almost, because it turns out that we have
$$
S_\pm = \Lambda^{ev,od}T^*\otimes\sqrt{\Lambda^n T},
$$
so in order to identify spinors with differential forms we have to choose a nowhere vanishing $n$--vector on $M$. The choices for this are parametrised by sections in $GL(n)/SL(n)$, that is, by a scalar function $e^{-\phi}$, $\phi\in C^\infty(M)$. This scalar field can be identified with the well--known dilaton field in string theory.
The Clifford action of elements $x\oplus\xi\in T\oplus T^*$ on spinors $\rho\in\Gamma(S_\pm)$ is given by
$$
(x\oplus\xi)\bullet\rho=-x\llcorner\rho+\xi\wedge\rho.
$$
For two spinors $\rho,\sigma\in\Gamma(S_\pm)$, we can define an invariant bilinear form (an inner product) as
$$
\langle\rho,\sigma\rangle=[\rho\wedge\hat{\sigma}]^n,
$$
where $\hat{\rho}=\pm\rho$ is the anti--automorphism of the Clifford algebra that gives a plus sign for $p$--forms with $p\equiv 0,3\mod 4$ and a minus sign otherwise.

To finally define $\GG$--structures, we look at specific subgroups of the generalised structure defined by $E=T\oplus T^*$, namely those which induce an additional metric $g$ (besides the natural inner product that exists on $E$) and a two--form $B$ on $T$. The pair $(g,B)$ is sometimes also called a generalised Riemannian metric.
The associated $H$--flux of the generalised Riemannian metric is then $H=dB$. One can also incorporate closed, but not globally exact $H$--fluxes by using gerbes, but we will not need that here. The existence of the datum $(g,B)$ reduces the structure group to $SO(n)\times SO(n)$. A {\em generalised $G$--structure} is a further reduction of this group to $G\times G$. We can describe this reduction in terms of the stabilised objects we obtain. In our case each $G$ preserves, as a subgroup of $Spin(n)$, a chiral spinor $\Psi_{1,2}$, and the invariant $\spin(n,n)$--spinor is $\rho=e^B[\Psi_1\otimes\Psi_2]$.

\section{Calibrations}\label{secCal}
To describe calibrated submanifolds on spaces with a generalised structure as introduced in the last section we have to extend our usual notion of a subspace. Before doing so, let us briefly recall the definition of calibrated subspaces in the classical case~\cite{hala82}. A $k$--form $\rho$ on an oriented vector space $W$ is called a calibration, if $\rho|_U\le\vol_U$ for some $k$--subplane $U$, where $\vol_U$ is the volume form induced by the Riemannian metric on $W$. The equality should hold for at least one subplane, which is said to be \emph{calibrated} with respect to $\rho$.

For generalised geometries, we have to consider not only a subspace $U$, but rather a pair $(U,F)$ with $F\in\Lambda^2U^*$. Later on, this $2$--form will account for a possible abelian\footnote{The concept presented here needs extension in order to incorporate nonabelian gauge groups.} gauge field on the D--brane. We can associate a so--called {\em pure} spinor to the datum $(U,F)$, namely the spinor annihilated by the isotropic subspace $e^{F}(U\oplus N^*U)$. Purity refers to the fact that the annihilator is of maximal dimension, namely $n$. This spinor is given, up to a multiple, by $e^F\bullet\widehat{\star\vol_U}$. Normalising, we define
$$
\rho_{(U,F)}:=\frac{e^F\bullet\widehat{\star\vol_U}}{\eunorm{e^F\bullet(\star\vol_U)}}.
$$
In analogy to the classical case, we speak of a calibration form $\rho^{ev,od}\in\Lambda^{ev,od}$ if
$$
\langle\rho,\rho_{(U,F)}\rangle\le 1
$$
for any $\rho_{U,F}$ and if the bound is met for at least one "`generalised"' subplane $(U,F)$. The calibration condition can also be cast in a slightly more familiar form for physicists (see Prop. 2.71 in~\cite{gmwi06}), namely
$$
[e^F\wedge\rho|_U]^k\le\sqrt{\det\left((g+B)|_U-F\right)}\vol_U.
$$
Everything we have done so far for vector spaces extends globally in an obvious way, for instance the calibration condition. Describing D--branes in string compactifications requires the reduction of the structure group not only on the topological level, but also imposes some integrability condition in analogy to the geometrical reductions in the classical case. In particular, we want conditions for the branes to preserve some part of the original supersymmetry which is equivalent to demanding that the corresponding submanifolds are volume minimising within their homology class.
In the presence of $F$--fields or fluxes, the volume functional gets replaced by the Born--Infeld functional containing contributions from the additional datum.

As has been proven in~\cite{gmwi06}, every calibrated submanifold will minimise the functional
$$
\int_U e^{-\phi}\sqrt{\det\left((g+B)|U-F\right)}
$$
if the calibration form is closed. To include the RR--potentials $C=\sum_k C^{(k)}$ of type II string theory into this setup amounts to considering an additional term in the functional and changing the integrability condition.

In the most general case we consider the integrability condition
$$
(d+H)\wedge e^{-\phi}\rho=(d+H)\wedge(e^B\wedge C),
$$
under which calibrated submanifolds minimise the functional,
$$
\int_U \left(e^{-\phi}\sqrt{\det\left((g+B)|_U-F\right)}-e^{B-F}\wedge C\right).
$$

\section{Example}
To make contact with calibrated subspaces used in type II string compactifications, let us check the general condition for the case of $n=6$ and $G=SU(3)$. An $\GxG{SU(3)}$--structure stabilises two $\spin(6,6)$--spinors $\rho^{od,ev}$, which can be decomposed into the tensor product of two chiral $\spin(6)$ spinors $\Psi_{L/R}$ as at the end of Section~\ref{secGG}. In the limiting case where the two spinors are equal, we obtain the well--known case of a classical $SU(3)$--structure.

We can calibrate with respect to $\Re\rho^{ev}=1-\omega^2/2$ or $\Re\rho^{od}=\Re\Omega$, where $\omega$ is the K\"ahler form and $\Omega$ the holomorphic $(3,0)$--form. For simplicity let $B=F=0$. The two respective calibration conditions then describe precisely the holomorphic cycles of B--branes and the Lagrangian cycles of A--branes. For non--zero $F$, we obtain an additional solution which corresponds to coisotropic A--branes (for a more detailed description of this and other examples, see~\cite{gmwi06}).

For a general $\GxG{SU(3)}$--structure, submanifolds of any even or odd dimension (depending on the choice of the calibration form) can occur. In particular, we could obtain isotropic A--branes of dimension one.

\begin{acknowledgement}
F. G. would like to thank the organisers of the RTN network conference in Naples for creating such a nice meeting. The work of F.~G. is supported by the Foundation for Fundamental Research of Matter (FOM) and the National Organisation for Scientific Research (NWO). The work of~F.~W. is supported by the SFB 647 of the German Research Council (DFG).
\end{acknowledgement}

\end{document}